# Towards ontology based BPMN implementation

CHHUN Sophea, MOALLA Néjib and OUZROUT Yacine

University of Lumiere Lyon2, laboratory DISP, France

**Natural language is understandable by human and not machine. None technical persons can only use natural language to specify their business requirements. However, the current version of Business process management and notation (BPMN) tools do not allow business analysts to implement their business processes without having technical skills. BPMN tool is a tool that allows users to design and implement the business processes by connecting different business tasks and rules together. The tools do not provide automatic implementation of business tasks from users' specifications in natural language (NL). Therefore, this research aims to propose a framework to automatically implement the business processes that are expressed in NL requirements. Ontology is used as a mechanism to solve this problem by comparing between users' requirements and web services' descriptions. Web service is a software module that performs a specific task and ontology is a concept that defines the relationships between different terms.**

## I. INTRODUCTION

This research focuses on the companies that use service oriented architecture (SOA) to deploy their business applications. SOA is an approach that provides guideline and technologies to build the applications in service-oriented world [18]. It provides the interoperability ability (platform independent) between different applications by using web services to exchange the resources. "Web service is a software system designed to support interoperable machine-to-machine interaction over a network" (W3C). Moreover, web service is described by using WSDL (Web Service Description Language) [13] in XML (eXtensible Markup Language) format. The SOA architecture allows the integration between web service applications and non web service applications without modifying the non web service applications. However, the programmers have to create an XML communication interface between the two applications.

Usually, the companies change their business requirements, and adopt new software applications, to react to the market evolutions and customers' needs. That causes the amount of web services increase and as well as the complexity of web services maintenance and management. Therefore, this research aims to propose the framework to reuse the existing web services when creating new business applications or re-engineering the existing business processes. Reusing the existing web services helps companies to save not only money but also time. However, it is possible that the existing web services cannot answer all users' needs. Therefore, this research has to propose the services composition or the service extension capability method to create a new web service that can answer the user's requirement.

Moreover, the business processes are implemented by using BPMN tools such as websphere or jdeveloper (oracle BPM suite). There are 2 steps to implement business process with BPMN tools, design and implementation. First, the business analyst designs the business processes by connecting different business tasks and rules together by using different connector types. Then, the programmer codes each task of the process and after that the users can deploy their business applications. So, it requires technical knowledge to implement the business processes that is not the case for business analysts. For this reason, this research aims to remove this technical constraint by creating a framework which can implement the business process automatically. The analysts design the business process and specify the description of each task. The framework will implement the process for them. In order to do this, we face some challenges problems to solve. First, the format to represent user's requirements: it should be simple and easy to express for non-technical person. Second, the matching between user's requirements and service functionalities to select the suitable service to perform the task. Third, the method to compose different web services together to create new services. The last one is the reconstruction of the business process with the implementation code.

This research will use ontology to define the relation between different terms in the same domain in a hierarchical way. It allows us to know the semantic meaning between different terms. In addition, there are 2 standards proposed by W3C to represent ontology, Resource Description Framework (RDF) [4] and Ontology Web Language (OWL) [5]. The detail is described in the next sections of this paper.

## II. RELATED WORKS

In SOA, there are 3 main components, service requesters, service providers and services registry [1]. The name of each component identifies clearly their functionalities. Service requesters are software modules that require some services to answer to their requirements. The service providers propose solutions or resources to the service requesters. All services in the company are stored in a services registry for future use. One example of service registries is universal description, discovery and integration (UDDI) [3].

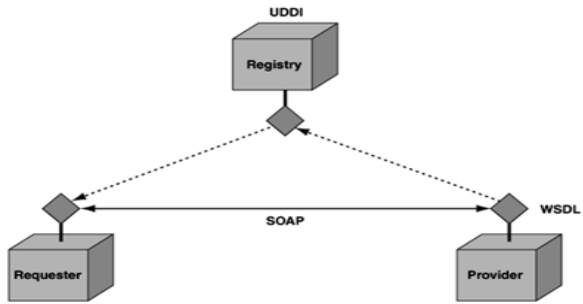

Fig. 1. Basic web service architecture [1]

There are two choices for implementing web services applications, orchestration and choreography [1-2]. In orchestration, an orchestra (a software module) manages the system process and tells to the service about which service that it has to communicate. However, in choreography there is no leader, every services are equal and they know which service they should work with. Furthermore, there are dynamic orchestration, static orchestration and semi-dynamic orchestration. In dynamic orchestration, the suitable web services are assigned to each task of the process at the run time. It is opposite to the static orchestration that it is done at the design time; web services are attached manually to each task by the programmer. Semi-dynamic orchestration combines the two ways together. This research study focuses on dynamic orchestration of web services when implementing business processes.

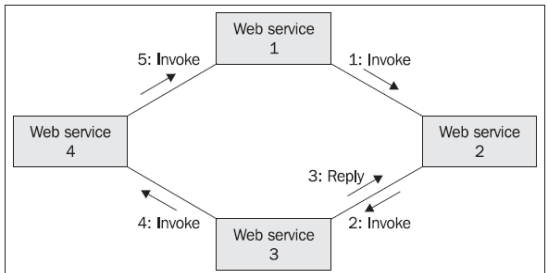

Fig. 2. Orchestration approach [2]

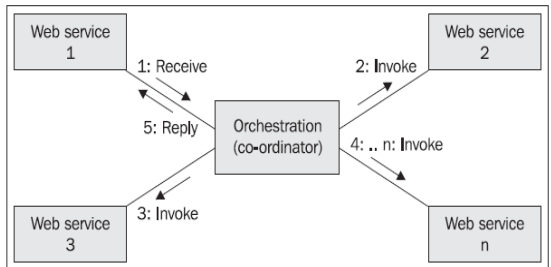

Fig. 3. Choreography approach [2]

The web service is described in XML (eXtensible Markup Language) format, XML is chosen to support interoperability between different parties who participate in the global system. Furthermore, the web service is described by using WSDL (Web Service Description Language) [13] and SOAP (Service Oriented Architecture Protocol) is used as a communication protocol between web services. Each web service has description and can be represented in ontology for the purpose of semantic matching when performing service selection. So, the ontology is built from WS (Web Service) description to store all concepts about web services including its domain. Then from users' requirements, the keywords are extracted to compare with ontology tree to obtain a suitable web service. In addition, the capabilities representation of WS in the ontology can be done in explicit, implicit and hybrid way [14]. In explicit representation, a pre-defined ontology is created to represent the concepts of all web services. However in implicit representation, the ontology is created in real time from the WS description and Hybrid is combined between explicit and implicit representation. Different ontology languages have been proposed to support machine readable and semantic matching such as OWL-S (Web Ontology Language-Semantic), WSMO (Web Service Modeling Ontology) [21],[23], CoOL (Context Ontology Language) and DAML+OIL (DARPA Agent Markup Language + Ontology Interchange Language). OWL-S is actually the supersede of DAML+OIL and has top level concept as service and contains three subontologies, service profile, service model and service grounding.

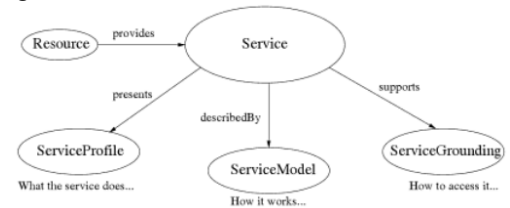

Fig. 4. OWL-S concept model [20] (p.120)

DAML-S is proposed by [19], it is not only the ontology representation but it is also a language to describe a WS, it provides semantic matching capability. Moreover, the authors propose DAML-S/UDDI matchmaker module that is used as an intermediate interface between users' requirements and the concepts that are stored in the ontology.

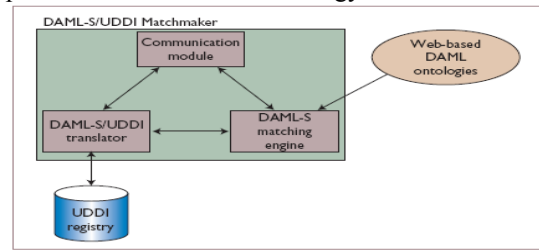

Fig. 5. DAML/UDDI matchmaker

WSMO [23] is created by using WSML (Web Service Modeling Language) and defined in four concepts, web services, goals, ontologies and mediators. Web service is specified by web service's capabilities and a set of interfaces that describe how to interact with web service. Goal is described by postcondition and effect, it describes the user's desire when requesting the web service. Ontology stores the terminologies accepted by web service and specifies the domain concept. The last one is mediator that specifies the mechanisms to allow web services to work together.

In addition, to represent the description of business processes, a specification is proposed in [8] called semantic of business vocabularies and business rules (SBVR). This specification is created for the machine readable purpose. Different languages (RuleSpeak, English structured language, Rabbit) and tools (ROO [16], ACEview) are created to manipulate this business process specification and build ontologies. However, this representation requires users to learn a new language that is opposite to this research objective, non technical persons can deploy business processes. Therefore, this research will use text to represent users' requirements.

[6] Propose a method to build ontology from the text description of a web service that is defined in javadoc but not from the WSDL file. Moreover, the authors make a comparison of two keywords extraction methods, POS and Deeper dependency linguistic analysis. These two methods are used to define the part of speech of each word in a sentence. Then, the JAPE rule is used to extract the keywords, for example only verb and noun phrase are extracted. Finally, they found that Deeper Dependency Linguistic Analysis provides better output ontology. On the other hand, [12] are not only propose a solution of the automatic building business process but they also use QoS (quality of service) to select the best web services in case many services are returned.

There are three ways to perform web service composition, static web services composition, dynamic web services composition and semi-dynamic web services composition. Some languages have proposed to perform a static service composition such as BPEL4WS (business process execution language for web service) [10], WSCI (Web Service Composition Interface) [11] and WS-BPEL (Web Service-Business Process Execution Language) [1]-[2]. Dynamic web service composition proposed by [9] and [12]. In [9], the authors use directed graph to perform the dynamic web service composition. It is done by comparing the output of one web service to the input of another web service. Moreover, the users provide the input and output values that are used as constraints to validate in service selection process. The authors also use the shortest path dynamic algorithm (Bellman Ford's algorithm) to select the best service composition path. [12] propose a goal oriented to perform dynamic services composition by extracting keywords from the goal description; then compare the keywords with ontology concept to create the final business process. In addition, [15] propose a semantic framework called OWL-T, T means task and OWL means ontology web language. This work uses different ontologies such as task, domain, process and service ontology to create business process. They propose an interesting task structure and as well as task type's hierarchy.

In [5], the authors propose BPMN 2.0 ontology which can be used as a knowledge base to understand about BPMN and as a syntax checker to validate concrete BPMN models. It is also used to identify the contradictions defined in BPMN specification models.

## III. PROPOSED SOLUTION

### A. Architecture of the Framework

This research framework takes the BPMN design specification that is generated by BPMN tool as an input. This design specification does not contain the implementation code and it is specified in the XML document. It contains all the information about business tasks, their descriptions and the information of different connectors that are used to connect each task together. The framework outputs the implementation version of the input design specification of business process which is also specified in XML format. Moreover, this output can be imported to BPMN tool for the deployment purpose.

To reach the research objectives, different modules of the framework are created.

1. Orchestra module: it reads the XML input document and then it uses xml parser to extract the content in the XML document. Then, it uses GetTaskPurpose to extract the description of each task. Its task is not only read the input XML document but it also responsible for reconstructing the process.

2. GetTaskPurpose module: it extracts the description of each task of the processes and passes it to KeywordExtraction module.

3. KeywordExtraction module: it separates each word of the input text from each other and identifies the part of speech of them. After that, it applies the pre-defined rules to get the keywords. Then, it passes the keywords into SemanticMatching module.

4. SemanticMatching module: after getting the keywords, it takes the keywords to search in service ontology tree to find the compatible web services. Then, it passes the found services to ServiceSelection module.

5. ServiceSelection module: after receving the input services list, if it is just one web service it will return to orchestra module. If not, it will calculate the QoS of all web services and select the one with the higher value of QoS; then returns the result to orchestra module. In case, none of web service is found in the process, it will notify to orchestration to call to ServiceComposition module.

6. ServiceComposition module: it performs the services composition to create a new web service that can answer to the requirement. If it cannot create a new web service, it will generate a message to notify to the users. Its input is a list of keywords that describe a business task.

The whole construction process can be described by the algorithm below:

```
For each process task {
    get the task description;
    extract the keywords;
    select a suitable WS;
    if a WS is found
        return WS;
    else if many WS are found
        compare QoS value;
        select the best WS;
        return WS;
    else {no WS is found}
        compose different WS together;
        if a new WS is created
            return newWS; else
            return messageError; }
```

The figure below describes the proposed framework architecture.

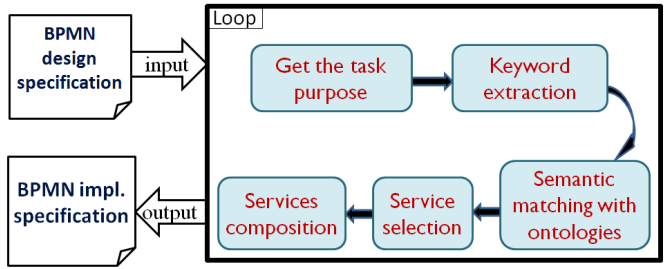

Fig. 6. Architecture of the framework

This framework is implemented using Jdeveloper and Oracle BPM suite.

### B. Input of the Framework

The input of the framework is an XML document which specifies the business process design by bpmn tool. This input is parsed to JDOM (Java XML Parser) parser and the parser will generate DOM tree. Then, a software module will be used to read the tasks' description from the tree.

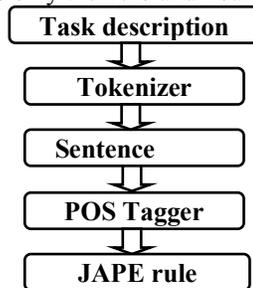

Fig. 7. Reading input XML file

### C. Keywords Extraction

This part extends an existing work of [6]. The framework uses POS tagger (Part Of Speech) that contains of Tokenizer, sentence splitter and POS tagger itself to analyze the linguistic of each word in a sentence. Tokenizer and sentence splitter modules are used to separate each keyword in a sentence from each others before passing to POS tagger to analyze part of speech of the work. Then, the framework uses JAPE rules [7] to extract keywords according to user's defined rules. This research extracts only the verb and noun phrase.

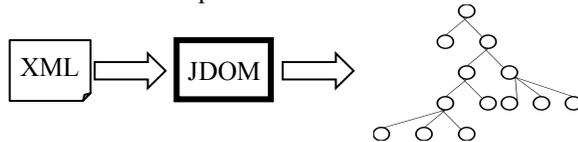

Fig. 8. Keywords extraction method

On the other hand, the process ontology is built from the previous successful implementation of service composition. It is created in order to avoid performing service composition process for the same requirements many times. So, if a service composition is needed, the framework will check in process repository. If the solution is exist, then take it. If not, perform services composition.

### D. Ontology Building

Web service and domain ontology are built by using the web service description of all web services that are stored in the repository. The noun phrase from WS description can be used to express the domain concept of the application. From the WS description, the keywords are extracted by the same method as described in session C. After getting the keywords, an algorithm is used to build domain and web service ontology. Some keywords might not necessarily to keep; therefore, the ontology pruning method is used to remove less frequency keywords compare to average value from the ontology trees. For the first time of ontology building, the domain experts will validate the ontology. Then Baseline method is used for automatic ontology pruning. This method assumes that the most frequent words dominate the domain of concepts and it removes low frequency keywords compare to total average of a keyword.

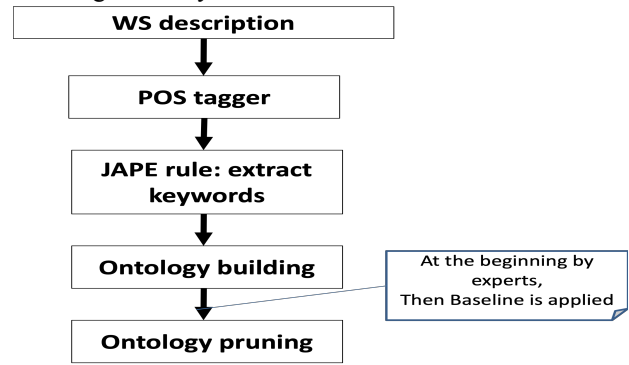

Fig. 9. WS and domain Ontology building

After studying different ontology language, DAML+OIL is selected because it is compatible with existing standards XML and RDFS (it inherits from RDFS) and easy to learn. Moreover, it is created to describe the structure of domain and it is object oriented (concepts as classes and roles as properties). DAML + OIL is also known as DAML-S and it combines the features of DAML and OIL.

### E. Web Service Selection

After performing service selection by comparing user's requirement and service description that performed by DAML-S/UDDI matchmaker, many matched web services might returns. Therefore, the QoS (Quality of Service) value is used to select the best suitable web service.

For the QoS's attributes, we decided to store the number of times a web service is available when a request is sent and its response time (the average response time value). The final attribute is the number of times a service has been called that expresses the service's popularity. These values are gotten from the log file that is generated by the weblogic when deploying the applications. In addition, the availability value of a web service stores the number of successful execution times of it. By comparing the value of the number of calls and the availability value, we know how good or bad a service is. The average execution duration of a web service is also used as a criteria for selecting a web service.

In short, a QoS express by availability value (a), average response time (b) and total number of called (n).

So, QoS is defined by

$$QoS = a - n \qquad (1)$$

In case, many services are returned, the algorithm below will be used:

For each WS$_i$

    ArrayA[i] = calculateQoS(WS$_i$);

    ArrayB = selectWSwithMaxQoSvalue(ArrayA);

    // only one WS that has maximum number of QoS

If numberOfelement(ArrayB)==1 then

    Return bestService;

Else

    Return WSwithMinAVGresponseTime;

### F. Web Service Structure

The web service structure is composed of 2 parts, general description and functional description. The general description is used to describe the general information about a web service.

In general description part of web service contains:

1). Publisher: to store the information of the WS's owner since in big companies, their applications might interconnect with other companies' applications (request services from other companies).

2). Component type: store the type of BPMN component that a service is created for. For example: human task, BPMN process, call service task.

3). LastUseDate: store the date when the web service is called.

4). URL: store the location of the web service.

5). QoS: to store some attributes that can be used to calculate the value of QoS. Those attributes are availability of service, its response time and number of calls (popularity).

6). Description: store description of a web service.

The functional description of web service contains:

1). OperationName: is the name of web service

2). Input: store the input object value of a web service

3). Output: store the output object value of a web service

The figure below shows the structure of the web service.

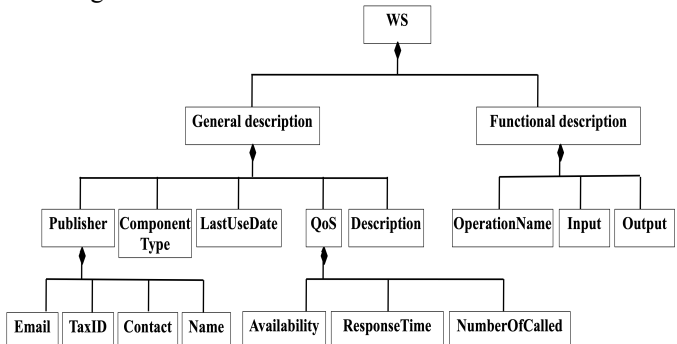

Fig. 10. Web service structure

## IV. CONCLUSION AND FUTURE WORK

This research study proposed a framework to automatically implement the business processes by reusing existing web services in the service repository. However, it still has many things to discover and improve. The current version of this framework can choose services to perform automatic implementation of business processes. However, if the existing web services cannot answer to users' requirements, we still don't know the best solution to perform dynamic web services composition yet. Web service composition method is in the process of discovering. We are also thinking about proposing the solution to extend the existing web service. If we compare web service to a class in object oriented programming, then it can be extended.

In addition, this research uses DAML+OIL to represent ontology of web services and QoS is used to select the most suitable web service to perform user's task.

Another future work is to use BPMN 2.0 ontology [9] to validate the output generated by the framework in order to make sure that all syntaxes are correct. We have to make sure also that the output can be exported to deploy in any BPMN tools.

## ACKNOWLEDGMENT

The authors gratefully acknowledge the support provided by the European Erasmus-Mundus Sustainable eTourism project 2010-2013.